\documentclass [useAMS]{mn2e}
\usepackage {graphicx}

\begin{document}

\title{A model of rotating hotspots for 3:2 frequency ratio of HFQPOs in black hole X-ray binaries}
%\date{Accepted }
%\pubyear{Accepted}
%\pagerange{\pageref{firstpage}--\pageref{endpage}}
%\onecolumn
\author[]{Ding-Xiong Wang$^{1,3}$, Yong-Chun Ye$^1$, Guo-Zheng Yao$^{1,2}$ and Ren-Yi Ma$^1$   \\
$1$ Department of Physics, Huazhong University of Science and Technology, Wuhan,430074,China \\
$2$ Department of Physics, Beijing Normal University, Beijing
100875, China \\
$3$ Send offprint requests to: D.-X. Wang (dxwang@hust.edu.cn) }
\maketitle

\begin{abstract}

We propose a model to explain a puzzling 3:2 frequency ratio of
high frequency quasi-periodic oscillations (HFQPOs) in black hole
(BH) X-ray binaries, GRO J1655-40, GRS 1915+105 and XTE J1550-564.
In our model a non-axisymmetric magnetic coupling (MC) of a
rotating black hole (BH) with its surrounding accretion disc
coexists with the Blandford-Znajek (BZ) process. The upper
frequency is fitted by a rotating hotspot near the inner edge of
the disc, which is produced by the energy transferred from the BH
to the disc, and the lower frequency is fitted by another rotating
hotspot somewhere away from the inner edge of the disc, which
arises from the screw instability of the magnetic field on the
disc. It turns out that the 3:2 frequency ratio of HFQPOs in these
X-ray binaries could be well fitted to the observational data with
a much narrower range of the BH spin. In addition, the spectral
properties of HFQPOs are discussed. The correlation of HFQPOs with
jets from microquasars is contained naturally in our model.

\end{abstract}


%-------------------------------- sec 1 --------------------------

\section{INTRODUCTION}

Quasi-periodic oscillations in X-ray binaries have become a very
active research field since the launch of the \textit{Rossi}
\textit{X-Ray Timing Explorer} (\textit{RXTE}; Bradt, Rothschild
{\&} Swank 1993). A key feature in these sources is that some of
high frequency quasi-periodic oscillations (HFQPOs) appear in
pairs. Five black hole (BH) X-ray binaries exhibit transient
HFQPOs, of which three sources have pairs occurring in GRO
J1655-40 (450, 300Hz; Strohmayer 2001a, hereafter S01a; Remillard
et al 1999), GRS 1915+105 (168, 113Hz; McClintock {\&} Remillard
2003, hereafter MR03), and XTE J1550-564 (276, 184Hz; Miller 2001;
Remillard et al 2002) with a puzzling 3:2 ratio of the upper
frequency to the lower frequency. Very recently, the 3:2 frequency
ratio (henceforth 3:2 ratio) was discovered in the X-ray outburst
of H1743-322 (240, 160Hz; Homan et al. 2004; Remillard et al.
2004). These discoveries give the exciting prospect of determining
BH mass and spin, as well as testing general relativity in the
strong-field regime.

A number of different mechanisms have been proposed to explain the
origin of HFQPO pairs in BH X-ray binaries. Strohmayer (S01a;
2001b) investigated combinations of the azimuthal and radial
coordinate frequencies in general relativity to explain the HFQPO
pairs in GRO J1655-40 and GRS 1915+105. Wagoner et al. (2001)
regarded the HFQPO pairs as fundamental g-mode and c-mode
discoseismic oscillations in a relativistic accretion disc.
Abramowicz {\&} Kluzniak (2001) explained the pairs in GRO
J1655-40 as a resonance between orbital and epicyclic motion of
accreting matter. It seems that more than one physical model is
required to fit all of the HFQPO observations.

As is well known, the Blandford-Znajek (BZ) process is an
effective mechanism for powering jets from AGNs and quasars, and
it is also applicable to jet production in microquasars (Blandford
{\&} Znajek 1977, Mirabel {\&} Rodrigues 1999). Recently, the
magnetic coupling (MC) of a rotating BH with its surrounding disc
has been investigated by some authors (Blandford 1999; Li 2000a,
2002a; Wang et al. 2002, hereafter W02), which can be regarded as
one of the variants of the BZ process. The MC process can be used
to explain a very steep emissivity in the inner region of the
disc, which is found by the recent \textit{XMM-Newton} observation
of the nearby bright Seyfert 1 galaxy MCG-6-30-15 (Wilms et al.
2001; Li 2002b; Wang et al. 2003a, hereafter W03a). In addition,
we explained the HFQPOs in X-ray binaries based on a model of a
rotating hotspot due to the MC of a rotating BH with the inner
region of the disc (Wang et al. 2003b, hereafter W03b). However,
the 3:2 ratio of HFQPOs has not been discussed by virtue of the MC
process.

Very recently, we discussed the condition for the coexistence of
the BZ and MC processes, and found that this coexistence always
accompanies the screw instability of the magnetic field in BH
magnetosphere, provided that the BH spin and the power-law index
for the variation of the magnetic field on the disc are greater
than some critical values (Wang et al. 2003c, 2004, hereafter W03c
and W04, respectively).

In this paper we propose a model to explain the 3:2 ratio of
HFQPOs in BH X-ray binaries by virtue of non-axisymmetric MC of a
rotating BH with its surrounding accretion disc. The upper
frequency arises from a rotating hotspot near the inner edge of
the disc, and the lower frequency is produced by the screw
instability somewhere away from the inner edge. It turns out that
the 3:2 ratio of HFQPOs in GRO J1655-40, GRS 1915+105 and XTE
J1550-564 is well fitted by energy transferred in the MC process,
where six parameters are used to describe the BH mass, spin and
distribution of a non-axisymmetric magnetic field. The 3:2 ratio
obtained in our model provides a much narrower range of the BH
spin compared with the other models. Furthermore, the jets from
these BH binaries are explained naturally by the BZ process
coexisting with the MC process.

This paper is organized as follows. In section 2 we give a brief
description of our model. In section 3 we fit the upper and lower
frequencies of HFQPOs by the inner and outer hotspots co-rotating
with the disc. The discussion is focused at the issues of the
outer hotspot. A scenario of how the presence of the screw
instability leads to the outer hotspot is given. In addition, the
time-scale of the screw instability is estimated by using an
equivalent R-L circuit, and the spectral properties of HFQPOs are
discussed by assuming the existence of corona above the disc. In
section 4 we discuss some issues of astrophysical meanings related
to the 3:2 ratio. Throughout this paper the geometric units $G = c
= 1$ are used.

%-------------------------------- sec 2 --------------------------

\section{DESCRIPTION OF OUR MODEL }

We intend to fit the 3:2 ratio of HFQPOs by virtue of an rotating
BH surrounded by a magnetized accretion disc based on the
following assumptions.

(\ref{eq1}) The configuration of the poloidal magnetic field is
shown in Fig. 1. The radius $r_{ms} $ indicates the inner edge of
the disc, being the innermost stable circular orbit (ISCO)
(Novikov {\&} Thorne 1973, hereafter NT73), and $r_{_{S}} $ is the
radius where the screw instability of the magnetic field might
occur, which is related to the angle $\theta _S $ by the same
mapping relation given in W04. As shown in Fig. 1, $\theta _S $ is
the angular boundary between the open and closed field lines on
the horizon, and $\theta _L $ is the lower boundary angle for the
closed field lines connecting the BH with the disc. Throughout
this paper $\theta _L = 0.45\pi $ is assumed in calculations.

%-------------------------------- f 1 --------------------------
\begin{figure}
\vspace{0.5cm}
\begin{center}
\includegraphics[width=8cm]{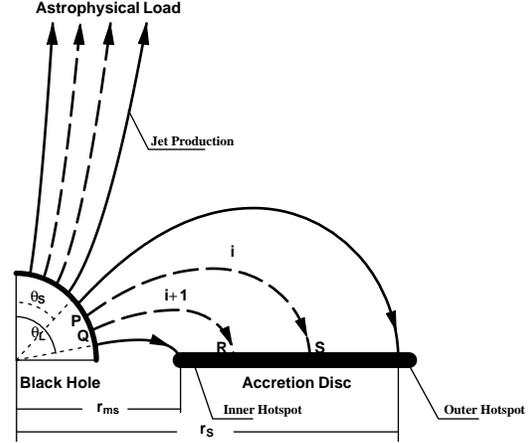}
\caption{Poloidal magnetic field connecting a rotating BH with a
remote astrophysical load and the surrounding disc.} \label{fig1}
\end{center}
\end{figure}

(\ref{eq2}) The poloidal magnetic field on the BH horizon is
non-axisymmetric as assumed in W03b, varying with the azimuthal
angle $\phi $ as follows (see Fig.2 in W03b),

\begin{equation}
\label{eq1} B_H^p \left( \phi \right) = \sqrt {\left\langle
{\left( {B_H^p } \right)^2} \right\rangle } f\left( \phi \right),
\quad f\left( \phi \right) \equiv \left\{ {\begin{array}{l}
 1 + \delta ,\mbox{ }0 < \phi < \Delta \phi , \\
 1,\mbox{ }\Delta \phi \le \phi \le 2\pi , \\
 \end{array}} \right.
\end{equation}

\noindent where $\sqrt {\left\langle {\left( {B_H^p } \right)^2}
\right\rangle } $ is root-mean-square of the poloidal magnetic
field over the angular coordinate from $\theta = 0$ to $\theta _L
$. The parameter $\delta $ is used to describe the strength of the
bulging magnetic field in the azimuthal region $0 < \phi < \Delta
\phi $.

(\ref{eq3}) The magnetosphere is assumed to be force-free outside
the BH and the disc, and the closed magnetic field lines are
frozen in the disc. The disc is thin and perfectly conducting,
lies in the equatorial plane of the BH with the inner boundary
being at ISCO.

(\ref{eq4}) The poloidal magnetic field varies as a power law with
the radial coordinate of the disc as follows,

\begin{equation}
\label{eq2} B_D^p \propto \xi ^{ - n},
\end{equation}

\noindent where $B_D^p $ represents the poloidal magnetic field on
the disc, and the parameter $n$ is the power-law index, and the
dimensionless radial coordinate is defined as $\xi \equiv r
\mathord{\left/ {\vphantom {r {r_{ms} }}} \right.
\kern-\nulldelimiterspace} {r_{ms} }$. The radius $r_{ms} $ is
related to the BH mass $M$ and the dimensionless spin parameter
$a_ * \equiv a \mathord{\left/ {\vphantom {a M}} \right.
\kern-\nulldelimiterspace} M$ as given in NT73.

(\ref{eq5}) The magnetic flux connecting the BH with its
surrounding disc takes precedence over that connecting the BH with
the remote load. As argued in W03c and W04, assumption 5 is
crucial for the coexistence of the BZ and MC processes.

The magnetosphere anchored in a Kerr BH and its surrounding disc
is described in Boyer-Lindquist coordinates, in which the
following parameters are involved (MacDonald and Thorne 1982,
hereafter MT82).

\begin{equation}
\label{eq3} \left\{ {\begin{array}{l}
 \Sigma ^2 = \left( {r^2 + a^2} \right)^2 - a^2\Delta \sin ^2\theta ,\mbox{
}\rho ^2 = r^2 + a^2\cos ^2\theta ,\mbox{ } \\
 \Delta = r^2 + a^2 - 2Mr,\mbox{ }\varpi = \left( {\Sigma \mathord{\left/
{\vphantom {\Sigma \rho }} \right. \kern-\nulldelimiterspace} \rho
}
\right)\sin \theta , \\
 \alpha = {\rho \sqrt \Delta } \mathord{\left/ {\vphantom {{\rho \sqrt
\Delta } \Sigma }} \right. \kern-\nulldelimiterspace} \Sigma . \\
 \end{array}} \right.
\end{equation}

In W03b the BZ and MC powers in non-axisymmetic case are related
to those in axisymmetic case by

\begin{equation}
\label{eq4} P_{BZ}^{NA} = \lambda P_{BZ}^A ,
\end{equation}

\begin{equation}
\label{eq5} P_{MC}^{NA} = \lambda P_{MC}^A ,
\end{equation}

\noindent where $P_{BZ}^{NA} $ and $P_{MC}^{NA} $ are the powers
in non-axisymmetic case, and $P_{BZ}^A $ and $P_{MC}^A $ are the
powers in axisymmetic case, respectively. The parameter $\lambda $
is expressed in terms of $\delta $ and $\varepsilon \equiv {\Delta
\phi } \mathord{\left/ {\vphantom {{\Delta \phi } {2\pi }}}
\right. \kern-\nulldelimiterspace} {2\pi }$ by

\begin{equation}
\label{eq6} \lambda = \left[ {\left( {1 + \delta }
\right)\varepsilon + \left( {1 - \varepsilon } \right)} \right]^2
= \left( {1 + \delta \varepsilon } \right)^2.
\end{equation}

Similarly, the BZ and MC torques in non-axisymmetic case are
related to those in axisymmetic case by

\begin{equation}
\label{eq7} T_{BZ}^{NA} = \lambda T_{BZ}^A ,
\end{equation}

\begin{equation}
\label{eq8} T_{MC}^{NA} = \lambda T_{MC}^A ,
\end{equation}

\noindent where $T_{BZ}^{NA} $ and $T_{MC}^{NA} $ are the torques
in non-axisymmetic case, and $T_{BZ}^A $ and $T_{MC}^A $ are the
torques in axisymmetic case, respectively. The powers $P_{BZ}^A $
and $P_{MC}^A $, and the torques $T_{BZ}^A $ and $T_{MC}^A $ are
given in W02 as follows.

\begin{equation}
\label{eq9} {P_{BZ}^A } \mathord{\left/ {\vphantom {{P_{BZ}^A }
{P_0 }}} \right. \kern-\nulldelimiterspace} {P_0 } = 2a_ * ^2
\int_0^\theta {\frac{k\left( {1 - k} \right)\sin ^3\theta d\theta
}{2 - \left( {1 - q} \right)\sin ^2\theta }} , \quad 0 < \theta <
\theta _S ,
\end{equation}

\begin{equation}
\label{eq10} {P_{MC}^A } \mathord{\left/ {\vphantom {{P_{MC}^A }
{P_0 }}} \right. \kern-\nulldelimiterspace} {P_0 } = 2a_ * ^2
\int_{\theta _S }^\theta {\frac{\beta \left( {1 - \beta }
\right)\sin ^3\theta d\theta }{2 - \left( {1 - q} \right)\sin
^2\theta }} , \quad \theta _S < \theta < \theta _L .
\end{equation}

%-------------------------------- eq 11 --------------------------

\begin{equation}
\label{eq11} \begin{array}{l} {T_{BZ}^A } \mathord{\left/
{\vphantom {{T_{BZ}^A } {T_0 }}} \right.
\kern-\nulldelimiterspace} {T_0 } = \\
\quad\quad 4a_ * \left( {1 + q} \right)\int_0^\theta {\frac{\left(
{1 - \beta } \right)\sin ^3\theta d\theta }{2 - \left( {1 - q}
\right)\sin ^2\theta }} , \quad 0 < \theta < \theta _S ,
\end{array}
\end{equation}

%-------------------------------- eq 12 --------------------------

\begin{equation}
\label{eq12} \begin{array}{l} {T_{MC}^A } \mathord{\left/
{\vphantom {{T_{MC}^A } {T_0 }}} \right.
\kern-\nulldelimiterspace} {T_0 } = \\
\quad\quad 4a_ * \left( {1 + q} \right)\int_{\theta _S }^\theta
{\frac{\left( {1 - \beta } \right)\sin ^3\theta d\theta }{2 -
\left( {1 - q} \right)\sin ^2\theta }} , \quad \theta _S < \theta
< \theta _L .
\end{array}
\end{equation}

\noindent In equations (\ref{eq9})---(\ref{eq12}), $k$ and $\beta
$ are the ratios of the angular velocities of the magnetic field
lines to the angular velocity of the BH horizon in the BZ and MC
processes, respectively. The quantity $q \equiv \sqrt {1 - a_ * ^2
} $ is a function of the BH spin, and $P_0 $ and $T_0 $ are
defined as

\begin{equation}
\label{eq13} \left\{ {\begin{array}{l}
 P_0 \equiv \left\langle {\left( {B_H^p } \right)^2} \right\rangle M^2
\approx B_4^2 m_{BH}^2 \times 6.59\times 10^{28}erg \cdot s^{ - 1}, \\
 T_0 \equiv \left\langle {\left( {B_H^p } \right)^2} \right\rangle M^3
\approx B_4^2 m_{BH}^3 \times 3.26\times 10^{23}g \cdot cm^2 \cdot
s^{ - 2},
\\
 \end{array}} \right.
\end{equation}

\noindent where $B_4 $ is the strength of the poloidal magnetic
field on the horizon in units of $10^4gauss$, and $m_{_{BH}} $ is
defined as $m_{BH} \equiv M \mathord{\left/ {\vphantom {M {M_
\odot }}} \right. \kern-\nulldelimiterspace} {M_ \odot }$.

So far six parameters are involved in our model, in which
$m_{_{BH}} $ and $a_
* $ are used for describing the mass and spin of the Kerr BH, and the three
parameters, $B_4 $, $\delta $ and $\varepsilon $ are used for
describing the non-axisymmetric magnetic field on the horizon, the
power-law index $n$ is for the variation of the magnetic field
with the radial coordinate of the disc.

%-------------------------------- sec 3 --------------------------

\section{INNER AND OUTER HOTSPOTS FOR 3:2 FREQUENCY RATIO}

As argued in W03b, the upper frequency of HFQPOs is modulated by
an inner hotspot, which corresponds to the maximum of function
$F_{QPO} $ as follows,

\begin{equation}
\label{eq14} F_{QPO} \equiv {r^2F} \mathord{\left/ {\vphantom
{{r^2F} {r_{ms}^2 }}} \right. \kern-\nulldelimiterspace} {r_{ms}^2
}F_0 = \xi ^2{\left( {F_{DA} + F_{MC} } \right)} \mathord{\left/
{\vphantom {{\left( {F_{DA} + F_{MC} } \right)} {F_0 }}} \right.
\kern-\nulldelimiterspace} {F_0 },
\end{equation}

\noindent where $F_{DA} $ is the radiation flux due to disc
accretion, and $F_{MC} $ is radiation flux arising from the energy
transferred from the rotating BH into the inner disc in the MC
process. We find that the radial coordinate of the inner hotspot,
$\xi _{\max } $, only depends on the parameters $a_ * $, and $ n$,
i.e., it is independent of the parameters $m_{_{BH}} $, $B_4 $,
$\delta $ and $\varepsilon $. Considering that the hotspot is
frozen at the disc, we have the upper frequency $\nu
_{QPO}^{upper} $ by substituting $\xi _{\max } $ into the
Keplerian angular velocity as follows,

\begin{equation}
\label{eq15} \nu _{QPO} = \nu _0 (\xi ^{3 \mathord{\left/
{\vphantom {3 2}} \right. \kern-\nulldelimiterspace} 2}\chi
_{ms}^3 + a_ * )^{ - 1},
\end{equation}

\noindent where $\nu _0 \equiv \left( {m_{BH} } \right)^{ -
1}\times 3.23\times 10^4Hz$. It turns out that $\nu _{QPO}^{upper}
$ depends on the three parameters, $m_{_{BH}} $, $a_ * $ and $n$.

The lower frequency of HFQPOs is fitted by an outer hotspot
rotating with the Keplerian angular velocity expressed by equation
(\ref{eq15}), and a scenario for the production of the outer
hotspot is given as follows.

It is well known that the magnetic field configurations with both
poloidal and toroidal components can be screw instable (Kadomtsev
1966; Bateman 1978). According to the Kruskal-Shafranov criterion
(Kadomtsev 1966), the screw instability will occur, if the
toroidal magnetic field becomes so strong that the magnetic field
line turns around itself about once. Recently, some authors
discussed the screw instability in BH magnetosphere (Gruzinov
1999a; Li 2000b; Tomimatsu et el. 2001). In W04 we suggested that
the criterion for the screw instability in the MC process could be
expressed by

\begin{equation}
\label{eq16} F_{Screw} \left( {a_ * ;\xi ,n} \right) - L
\mathord{\left/ {\vphantom {L {\left( {2\pi \varpi _D } \right)}}}
\right. \kern-\nulldelimiterspace} {\left( {2\pi \varpi _D }
\right)} \le 0,
\end{equation}

\noindent where $L$ is the poloidal length of the closed field
line connecting the BH and the disc, $\varpi _{_{D}} $ is the
cylindrical radius on the disc, and $F_{Screw} \left( {a_ * ;\xi
,n} \right)$ is a function of $a_ * $, $\xi $ and $n$.

Setting equality in equation (\ref{eq16}), we obtain the critical
radial coordinate $\xi _S \equiv {r_S } \mathord{\left/ {\vphantom
{{r_S } {r_{ms} }}} \right. \kern-\nulldelimiterspace} {r_{ms} } $
for the screw instability, which depends on the parameters $a_ * $
and $n$.

It is believed that a disc is probably surrounded by a
high-temperature corona analogous to the solar corona (Liang {\&}
Price 1977; Haardt 1991; Zhang et al 2000). Very recently, some
authors argued that the coronal heating in some stars including
the Sun is probably related to dissipation of currents, and very
strong X-ray emissions arise from variation of magnetic fields
(Galsgaard {\&} Parnell 2004; Peter et al. 2004).

Analogously, if the corona exists above the disc in our model, we
expect that the corona above $\xi _S $ might be heated by the
induced current due to the screw instability of the
non-axisymmetric magnetic field. Therefore a very strong X-ray
emission would be produced above the rotating outer hotspot, and
the lower frequency $\nu _{QPO}^{lower} $ is obtained directly by
substituting $\xi _S $ into equation (\ref{eq15}).

The time-scale of the screw instability can be estimated by using
an equivalent circuit as shown in Fig. 2, in which the segments
\textbf{\textit{LM}} and \textbf{\textit{KN}} represent the two
adjacent magnetic surfaces connecting the BH horizon and the outer
hotspot as shown in Fig. 2a. Considering the existence of the
toroidal magnetic field threading each loop, we introduce an
inductor into the circuit, which is represented by the symbol
$\Delta L$ in Fig. 2b.

\begin{equation}
\label{eq17} \Delta \varepsilon _H = \left( {{\Delta \Psi ^p}
\mathord{\left/ {\vphantom {{\Delta \Psi ^p} {2\pi }}} \right.
\kern-\nulldelimiterspace} {2\pi }} \right)\Omega _H , \quad
\Delta \varepsilon _D = - \left( {{\Delta \Psi ^p} \mathord{\left/
{\vphantom {{\Delta \Psi ^p} {2\pi }}} \right.
\kern-\nulldelimiterspace} {2\pi }} \right)\Omega _D ,
\end{equation}

\noindent where $\Delta \Psi ^p$ is the flux of the poloidal
magnetic field sandwiched by the magnetic surfaces\textbf{\textit{
LM}} and \textbf{\textit{KN}}. The resistance at the BH horizon is
given by

\begin{equation}
\label{eq18} \Delta R_H = \sigma _H \frac{\Delta l}{2\pi \varpi _H
} = \frac{2\rho _H \Delta \theta _H }{\varpi _H },
\end{equation}

\noindent where $\sigma _H = 4\pi = 377ohm$ is the surface
resistivity of the BH horizon (MT82), and $\Delta \theta _H $ is
the angle on the horizon spanned by the two surfaces. The
quantities $\rho _{_{H}} $ and $\varpi _{_{H}} $ are the Kerr
metric parameters at the horizon. The inductance $\Delta L$ in the
R-L circuit is defined by

\begin{equation}
\label{eq19} \Delta L = {\Delta \Psi ^T} \mathord{\left/
{\vphantom {{\Delta \Psi ^T} {I^p}}} \right.
\kern-\nulldelimiterspace} {I^p},
\end{equation}

\noindent where $I^p$ and $\Delta \Psi ^T$ are the poloidal
current flowing in the circuit and the flux of the toroidal
magnetic field threading the circuit, respectively. The flux
$\Delta \Psi ^T$ can be integrated over the region \textit{KLMN}
as follows,

\begin{equation}
\label{eq20} \Delta \Psi ^T = \oint_{loop} {B^T\sqrt {g_{rr}
g_{\theta \theta } } dr} d\theta ,
\end{equation}

\noindent where the toroidal magnetic field measured by
``zero-angular-momentum observers'' (ZAMOs) is given as

\begin{equation}
\label{eq21} B^T = {2I^p} \mathord{\left/ {\vphantom {{2I^p}
{\left( {\alpha \varpi } \right)}}} \right.
\kern-\nulldelimiterspace} {\left( {\alpha \varpi } \right)},
\end{equation}

\noindent where $\alpha $ is the lapse function defined in
equation (\ref{eq3}).

Unfortunately, the geometric shapes of the magnetic surfaces
\textbf{\textit{LM}} and \textbf{\textit{KN}} are unknown. As an
approximate estimation we assume that the surfaces
\textbf{\textit{LM}} and \textbf{\textit{KN}} are formed by
rotating the two adjacent circles around the symmetric axis, and
$\Delta \Psi ^T$ can be calculated by integrating over the region
\textbf{\textit{KLMN}} as shown in Fig. 3. Finally, we calculate
the quantity $\tau \equiv {\Delta L} \mathord{\left/ {\vphantom
{{\Delta L} {\Delta R_H }}} \right. \kern-\nulldelimiterspace}
{\Delta R_H }$ by incorporating equations
(\ref{eq18})---(\ref{eq21}), which is independent of $I^P$.

%-------------------------------- f 2 --------------------------
\begin{figure}
\vspace{0.5cm}
\begin{center}
{\includegraphics[width=6cm]{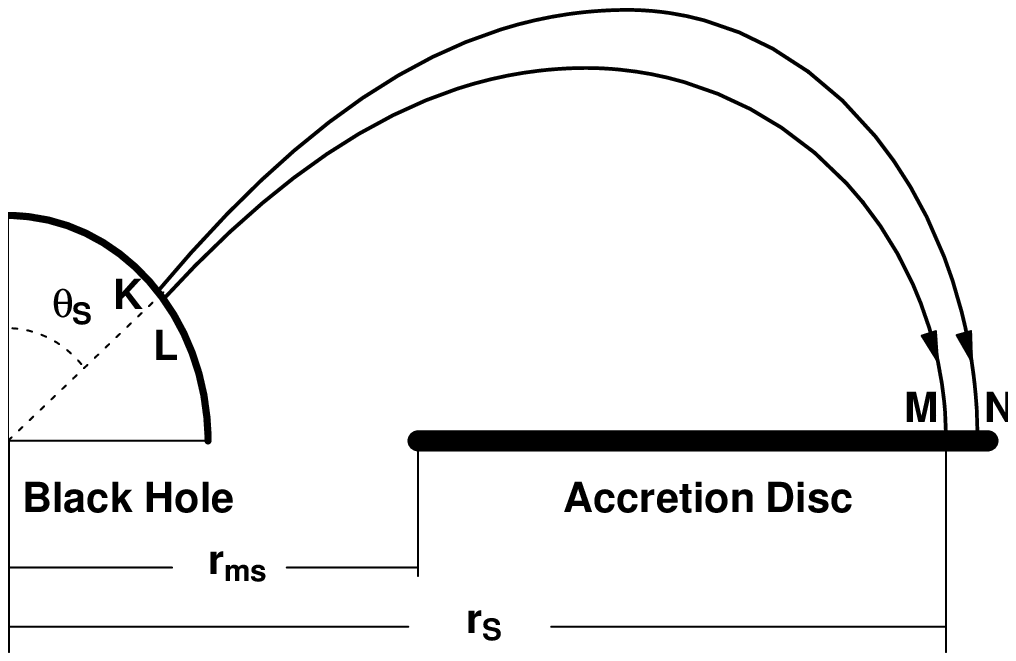}
 \centerline{(a)}
 \includegraphics[width=5cm]{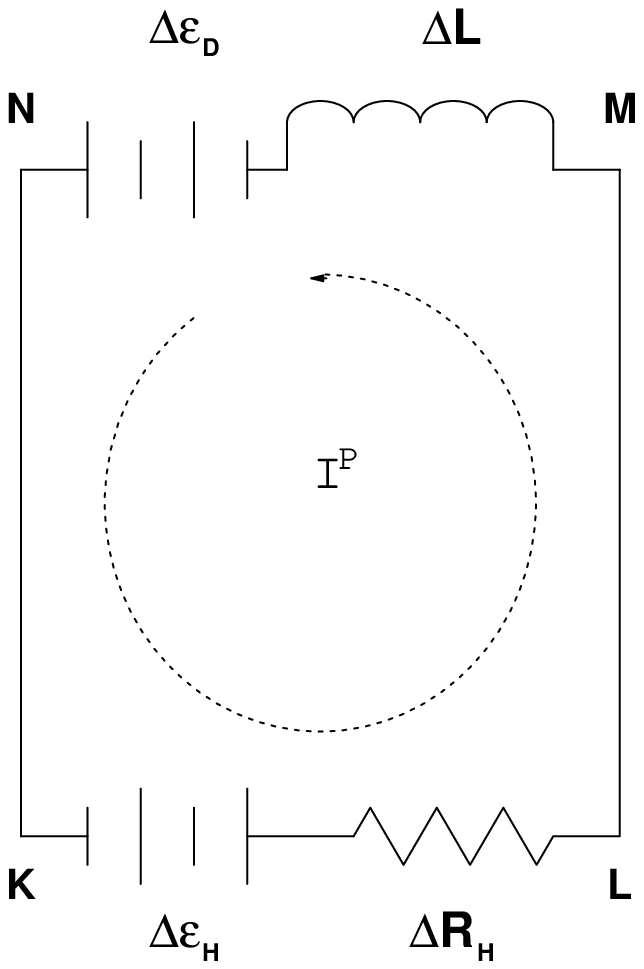}
 \centerline{(b)}}
 \caption{(a) Two adjacent magnetic surfaces connecting the BH horizon
  and the outer hotspot, (b) one loop of equivalent circuit for screw
  instability.}\label{fig2}
\end{center}
\end{figure}

%-------------------------------- f 3 --------------------------
\begin{figure}
\vspace{0.5cm}
\begin{center}
\includegraphics[width=8cm]{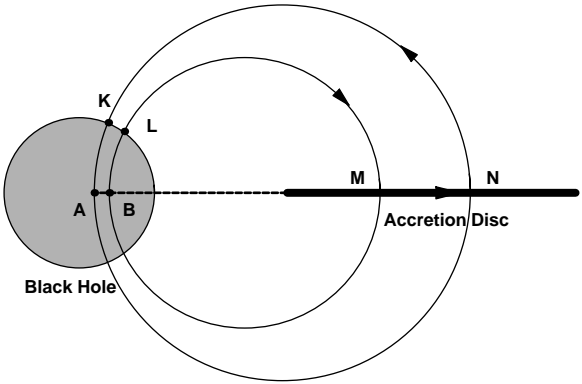}
\caption{Two adjacent magnetic surfaces produced by two rotating
circles, where the arrows represent the current flowing in the R-L
circuit.} \label{fig3}
\end{center}
\end{figure}

As a simple analysis we can divide one event of the screw
instability into two stages. In stage 1 the instability starts,
and the magnetic energy is released rapidly, the poloidal current
and thus the toroidal magnetic field decrease to zero in very
short time. In stage 2 the toroidal magnetic field recovers
gradually with the increasing poloidal current $I^P$ in the R-L
circuit, until the next event of the instability occurs according
to the criterion (\ref{eq16}). It seems reasonable that the
duration in stage 1 is much less than that in stage 2, and the
time-scale of the screw instability depends mainly on the duration
of stage 2, which is regarded as the relax time of the poloidal
current $I^P$ in the R-L circuit. The current $I^P$ is governed by
the following equation,

\begin{equation}
\label{eq22} \Delta L\frac{dI^P}{dt} + \Delta R_H I^P = \Delta
\varepsilon _H + \Delta \varepsilon _L .
\end{equation}

\noindent Setting the initial condition, $I^P = 0$, we have the
solution,

%-------------------------------- eq 23 --------------------------

\begin{equation}
\label{eq23} I^P\left( t \right) = I_{steady}^p \left( {1 - e^{ -
t \mathord{\left/ {\vphantom {t \tau }} \right.
\kern-\nulldelimiterspace} \tau }} \right),
\end{equation}

\noindent where $I_{steady}^p $ is the steady current in the R-L
circuit, and it reads

\begin{equation}
\label{eq24} I_{steady}^p = {\left( {\Delta \varepsilon _H +
\Delta \varepsilon _L } \right)} \mathord{\left/ {\vphantom
{{\left( {\Delta \varepsilon _H + \Delta \varepsilon _L } \right)}
{\Delta R_H }}} \right. \kern-\nulldelimiterspace} {\Delta R_H }.
\end{equation}

From equation (\ref{eq23}) we obtain that the poloidal current
attains 99.3{\%} of $I_{steady}^p $ in the relax time $t_{relax} =
5\tau $, implying the recovery of the toroidal magnetic field.
Thus the time-scale of the screw instability can be estimated as

\begin{equation}
\label{eq25} t_{Screw} > t_{relax} = 5\tau .
\end{equation}

Summarizing the above results, we have the upper and lower
frequencies of HFQPOs, and the corresponding values of $\xi _{\max
} $, $\xi _S $ and $t_{relax} $ as shown in Table 1. It is easy to
check from Table 1 that the time-scale of the screw instability is
generally greater than the corresponding periods of HFQPOs, i.e.,

\begin{equation}
\label{eq26} t_{Screw} > t_{relax} > 1 \mathord{\left/ {\vphantom
{1 {\nu _{QPO}^{upper}
> 1 \mathord{\left/ {\vphantom {1 {\nu _{QPO}^{lower} }}} \right.
\kern-\nulldelimiterspace} {\nu _{QPO}^{lower} }}}} \right.
\kern-\nulldelimiterspace} {\nu _{QPO}^{upper} > 1 \mathord{\left/
{\vphantom {1 {\nu _{QPO}^{lower} }}} \right.
\kern-\nulldelimiterspace} {\nu _{QPO}^{lower} }}.
\end{equation}

As pointed out in MR03, the spectral properties of accreting black
holes in states which show HFQPOs are usually dominated by a steep
power-law (SPL) component. However, the origin of X-ray power-law
remains controversial. Most of models for the SPL state invoke
inverse Compton scattering as the operant radiation mechanism, and
the radiation spectrum formed through Comptonization of
low-frequency photons in a hot thermal plasma cloud may be
described by a power law (Pozdnyakov, Sobol {\&} Syunyaev 1983).
It is noted that the origin of the Comptonizing electrons involves
magnetic instabilities in the accretion disc (Poutanen {\&} Fabian
1999), with which the screw instability in our model is
consistent.

%-------------------------------- tab 1 --------------------------
\begin{table*}
\caption{The 3:2 ratio of HFQPOs produced by the inner and outer
hotspots with $\delta = 0.5$ and $\varepsilon = 0.2$. Henceforth
BHB-I, II and III represent GRO J1655-40, GRS 1915+105 and XTE
J1550-564, respectively.}
\begin{tabular}
{|p{25pt}|p{25pt}|p{20pt}|p{23pt}|p{49pt}|p{26pt}|p{54pt}|p{26pt}|p{26pt}|p{54pt}|p{26pt}|}
\hline \raisebox{-1.50ex}[0cm][0cm]{\textbf{BHB}}&
\raisebox{-1.50ex}[0cm][0cm]{$a_ * $}&
\raisebox{-1.50ex}[0cm][0cm]{$n$}&
\raisebox{-1.50ex}[0cm][0cm]{$m_{_{BH}} $ }&
\raisebox{-1.50ex}[0cm][0cm]{$t_{relax} $ (sec)}&
\multicolumn{3}{|p{135pt}|}{Inner Hotspot} &
\multicolumn{3}{|p{126pt}|}{Outer Hotspot}  \\
\cline{6-11}
 &
 &
 &
 &
 &
$\xi _{\max } $ & ${E_{HS}^{upper} } \mathord{\left/ {\vphantom
{{E_{HS}^{upper} } {B_4^{1 / 2} }}} \right.
\kern-\nulldelimiterspace} {B_4^{1 / 2} }$ \par \textit{(kev)} &
$\nu _{QPO}^{upper} $ \par \textit{(Hz)} & $\xi _{SC} $ &
${E_{HS}^{lower} } \mathord{\left/ {\vphantom {{E_{HS}^{lower} }
{B_4^{1 / 2} }}} \right. \kern-\nulldelimiterspace} {B_4^{1 / 2}
}$ \par \textit{(kev)} &
$\nu _{QPO}^{lower} $ \par \textit{(Hz)}  \\
\hline \raisebox{-1.50ex}[0cm][0cm]{\textbf{I}}& 0.777& 6.22& 6.8&
$7.61\times 10^{ - 3}$ & 1.277& $7.23\times 10^{ - 3}$ & 450&
2.015& $4.96\times 10^{ - 3}$ &
300 \\
\cline{2-11}
 &
0.730& 6.27& 5.8& $6.57\times 10^{ - 3}$ & 1.281& $6.72\times 10^{
- 3}$ & 450& 2.097& $4.46\times 10^{ - 3}$ &
300 \\
\hline \raisebox{-1.50ex}[0cm][0cm]{\textbf{II}}& 0.773& 6.23& 18&
$2.02\times 10^{ - 2}$ & 1.277& $7.19\times 10^{ - 3}$ & 168&
2.021& $4.92\times 10^{ - 3}$ &
113 \\
\cline{2-11}
 &
0.603& 6.17& 10& $1.11\times 10^{ - 2}$ & 1.297& $5.55\times 10^{
- 3}$ & 168& 2.428& $3.27\times 10^{ - 3}$ &
113 \\
\hline \raisebox{-1.50ex}[0cm][0cm]{\textbf{III}}& 0.788& 6.19&
11.5& $1.28\times 10^{ - 2}$ & 1.276& $7.37\times 10^{ - 3}$ &
276& 2.000& $5.09\times 10^{ - 3}$ &
184 \\
\cline{2-11}
 &
0.696& 6.29& 8.5& $9.72\times 10^{ - 3}$ & 1.284& $6.38\times 10^{
- 3}$ & 276& 2.159& $4.14\times 10^{ - 3}$ &
184 \\
\hline
\end{tabular}
\label{tab1}
\end{table*}
\begin{table*}
 %\centering
\begin{minipage}{170mm}
\textbf{Note:} The value ranges of the BH mass corresponding to
GRO J1655-40, GRS 1915+105 and XTE J1550-564 are adopted from
Greene et al. (2001), MR03 and Orosz et al. (2002), respectively.
 \end{minipage}
\end{table*}

Suppose that the spectral properties of the inner and outer
hotspots arise from blackbody spectra, the effective radiation
temperature $\left( {T_{HS} } \right)_{eff} $ is expressed by

\begin{equation}
\label{eq27} T_{eff} = T_0 \left[ {f_{DA} + f_{MC} } \right]^{1
\mathord{\left/ {\vphantom {1 4}} \right.
\kern-\nulldelimiterspace} 4},
\end{equation}

\begin{equation}
\label{eq28} T_0 = \left( {{F_0 } \mathord{\left/ {\vphantom {{F_0
} \sigma }} \right. \kern-\nulldelimiterspace} \sigma } \right)^{1
\mathord{\left/ {\vphantom {1 4}} \right.
\kern-\nulldelimiterspace} 4} \approx 4.8\times 10^5B_4^{1
\mathord{\left/ {\vphantom {1 2}} \right.
\kern-\nulldelimiterspace} 2} \left( K \right),
\end{equation}

\noindent where $\sigma $ is the Stefan-Boltzmann constant. The
energy of the inner and outer hotspots can be estimated by

\begin{equation}
\label{eq29} E_{HS} \equiv k_B T_{eff} = E_0 \left[ {f_{DA} +
f_{MC} } \right]^{1 \mathord{\left/ {\vphantom {1 4}} \right.
\kern-\nulldelimiterspace} 4},
\end{equation}

\begin{equation}
\label{eq30} E_0 = k_B \left( {{F_0 } \mathord{\left/ {\vphantom
{{F_0 } \sigma }} \right. \kern-\nulldelimiterspace} \sigma }
\right)^{1 \mathord{\left/ {\vphantom {1 4}} \right.
\kern-\nulldelimiterspace} 4} \approx 4.14\times 10^{ - 2}B_4^{1
\mathord{\left/ {\vphantom {1 2}} \right.
\kern-\nulldelimiterspace} 2} \left( {kev} \right),
\end{equation}

\noindent where $k_B $ is the Boltzmann constant. Substituting
$\xi _{\max } $ and $\xi _S $ into equation (\ref{eq29}), we have
the energy of the inner and outer hotspots $E_{HS}^{upper} $ and
$E_{HS}^{lower} $ as shown in Table 1.

Inspecting equation (\ref{eq29}) and the expressions for $f_{DA} $
and $f_{MC} $ given in W03b, we find that both $E_{HS}^{upper} $
and $E_{HS}^{lower} $ depend not only on the parameters, $a_ * $
and $n$, but also on the parameters for the non-axisymmetric
magnetic field, $B_4^{1 / 2} $, $\delta $ and $\varepsilon $. It
is worthy to point out that both $E_{HS}^{upper} $ and
$E_{HS}^{lower} $ are independent of the BH mass of the X-ray
binaries.

%-------------------------------- sec 4 --------------------------

\section{DISCUSSION}

Inspecting Table 1, we find that the 3:2 ratios of HFQPOs in GRO
J1655-40, GRS 1915+105 and XTE J1550-564 are well fitted by
adjusting the two parameters, $a_ * $ and $n$, for the given value
ranges of $m_{_{BH}} $. In this section we discuss some issues of
astrophysical meanings related to the 3:2 ratio of HFQPOs in the
BH binaries.

%-------------------------------- sub 1 --------------------------

\subsection{3:2 ratio and magnetic field on the BH horizon}

It is pointed out that the twin HFQPOs are not always detected
together in GRO J1655-40, i.e., 300 Hz QPO is observed sometimes
without 450 Hz detection, while 450 Hz QPO is observed sometimes
without 300 Hz detection (S01a). In our model the upper and lower
frequencies of HFQPOs are produced by the inner and outer
hotspots, which are located at different sites of the disc. In
addition, our model is proposed based on the non-axisymmetric
magnetic field on the horizon, and two different physical
mechanisms are involved, i.e., the MC process for the inner
hotspot and the screw instability for the outer hotspot. Since the
non-axisymmetric magnetic field cannot be stationary on the
horizon, we can understand that the twin HFQPOs do not have to
occur simultaneously in some cases.

In Table 1, both $E_{HS}^{upper} $ and $E_{HS}^{lower} $ reach the
energy level of emitting X-ray, provided that the root-mean square
of the magnetic field on the horizon is strong enough to be $ \sim
10^9gauss$ ($B_4 \sim 10^5)$. Considering that a system cannot
radiate a given flux at less than the blackbody temperature (Frank
et al. 1992), the strength of magnetic field of $ \sim 10^9gauss$
might be regarded as a upper limit for energy level of emitting
X-ray in BH binaries.

It is pointed out in S01a that the energy spectra of the 300 Hz
QPO appear to be significantly different from those of the 450 Hz
QPO in GRO J1655-40: The former has a typical amplitude in the
2--12 keV band, while the latter is detected in the hard band,
13--27 keV. It is also shown in Fig. 4.16 of MR03 that the power
densities corresponding to the upper frequency are greater than
those corresponding to the lower frequency of HFQPOs in some BH
X-ray binaries, such as GRO J1655-40 (450, 300Hz) and XTE
J1550-564 (276, 184Hz). These observations are consistent with the
results obtained in our model, i.e., $E_{HS}^{upper} $ is greater
than $E_{HS}^{lower} $ for each BH binary as shown in Table 1.

%-------------------------------- sub 2 --------------------------

\subsection{BH spin and mass constrained by 3:2 ratio}

It is widely believed that HFQPOs in BH binaries might be a unique
timing signature constraining the BH mass and spin via a model
rooted in general relativity. Thus a precise measurement of the
3:2 ratio of HFQPOs becomes an approach to estimate the BH mass
and spin. Combining the fitting results in Table 1 with the
observations of the three BH binaries, we discuss the issues
related to the BH mass and spin as follows.

The BH spin in GRO J1655-40 has been estimated by some authors.
Interpreting the X-ray spectra, Zhang, Cui {\&} Chen (1997)
suggested that the most likely value is $0.7 < a_ * < 1$, while
Sobczak et al. (1999) give an upper limit $a_ * < 0.7$. Assuming
that a bright spot appears near the radius of the maximal proper
radiation flux from a disc around a rotating BH, Gruzinov (1999b)
inferred an upper limit $a_ * < 0.6$ for the BH in GRO J1655-40.
Abramowicz {\&} Kluzniak (2001) suggested that the 3:2 ratio in
GRO J1655-40 arises from a resonance between orbital and epicyclic
motion of accreting matter, and they estimated that the value
range of the BH spin is $0.2 < a_
* < 0.65$. Compared with the above estimation ranges, we provide a much
narrower range for the BH spin in GRO J1655-40, $0.730 < a_ * <
0.777$, as shown in Table 1, where the BH spins in GRS 1915+105
and XTE J1550-564 are also constrained in rather narrow ranges,
$0.603 < a_ * < 0.773$ and $0.696 < a_ * < 0.788$, respectively.

%-------------------------------- sub 3 --------------------------

\subsection{3:2 ratio and jets from microquasars}

The BH X-ray binaries, GRO J1655-40, GRS 1915+105 and XTE
J1550-564, are also regarded as microquasars, from which
relativistic jets are observed (Mirabel {\&} Rodrigues 1998, 1999
and references therein; see also Table 4.2 of MR03). This
correlation can be explained very naturally by virtue of the
magnetic field configuration as shown in Fig. 1. The jet power is
assumed to be provided by the BZ mechanism and expressed by
equations (\ref{eq4}) and (\ref{eq9}). Substituting $a_ * $, $n$,
$m_{_{BH}} $ and $\xi _S $ into these equations, we have the jet
powers as shown in Table 2. In calculations the parameter $k =
0.5$ is taken for the optimal BZ power (MT82).

%-------------------------------- tab 2 --------------------------
\begin{table*}
\caption{The jet powers accompanying 3:2 ratio of HFQPOs with
$\delta = 0.5$ and $\varepsilon = 0.2$.}
\begin{tabular}
{|p{32pt}|p{59pt}|p{59pt}|p{59pt}|p{59pt}|p{95pt}|} \hline BHB&
$a_ * $& $n$& $m_{_{BH}} $& $\theta _S $&
${P_{BZ}^{NA} } \mathord{\left/ {\vphantom {{P_{BZ}^{NA} } {B_4^2 }}} \right. \kern-\nulldelimiterspace} {B_4^2 }\left( {{erg} \mathord{\left/ {\vphantom {{erg} s}} \right. \kern-\nulldelimiterspace} s} \right)$ \\
\hline \raisebox{-1.50ex}[0cm][0cm]{I}& 0.777& 6.22& 6.8&
$50.7^0$&
$7.09\times 10^{28}$ \\
\cline{2-6}
 &
0.730& 6.27& 5.8& $48.9^0$&
$3.98\times 10^{28}$ \\
\hline \raisebox{-1.50ex}[0cm][0cm]{\textbf{II}}& 0.773& 6.23& 18&
$50.5^0$&
$4.87\times 10^{29}$ \\
\cline{2-6}
 &
0.603& 6.17& 10& $42.4^0$&
$4.68\times 10^{28}$ \\
\hline \raisebox{-1.50ex}[0cm][0cm]{\textbf{III}}& 0.788& 6.19&
11.5& $50.9^0$&
$2.14\times 10^{29}$ \\
\cline{2-6}
 &
0.696& 6.29& 8.5& $47.7^0$&
$7.01\times 10^{28}$ \\
\hline
\end{tabular}
\label{tab2}
\end{table*}

Mirabel {\&} Rodriguez (1999) pointed out that GRS 1915+105 may
have a short-term jet power of $ \sim \mbox{10}^{39}{erg}
\mathord{\left/ {\vphantom {{erg} s}} \right.
\kern-\nulldelimiterspace} s$ in its very high state, which is a
large fraction of the observed accretion power. From Table 2 we
have the angular boundary and the corresponding BZ power for GRS
1915+105 as follows.

\begin{equation}
\label{eq31} P_{BZ}^{NA} = 4.87\times 10^{29}B_4^2 \left( {{erg}
\mathord{\left/ {\vphantom {{erg} s}} \right.
\kern-\nulldelimiterspace} s} \right), \quad 0 < \theta < 50.5^0,
\end{equation}

\begin{equation}
\label{eq32} P_{BZ}^{NA} = 4.68\times 10^{28}B_4^2 \left( {{erg}
\mathord{\left/ {\vphantom {{erg} s}} \right.
\kern-\nulldelimiterspace} s} \right), \quad 0 < \theta _S <
42.4^0.
\end{equation}

Equations (\ref{eq31}) and (\ref{eq32}) imply that a large
fraction of the rotating energy is extracted from the BH contained
in GRS 1915+105 for the jet power, and $P_{BZ}^{NA} $ can reaches
$ \sim \mbox{10}^{39}{erg} \mathord{\left/ {\vphantom {{erg} s}}
\right. \kern-\nulldelimiterspace} s$, provided that the magnetic
field on the horizon is strong enough to attain $ \sim 10^9gauss$
($B_4 \sim 10^5)$. This strength of the magnetic field is the same
as that required by the inner and outer hotspots for emitting
X-ray. By using equations (\ref{eq4}) and (\ref{eq9}) we find that
the BZ power increases monotonically with the power-law index $ n$
for the given BH spin $a_ * $, while it varies non-monotonically
with $a_ * $ for the given $n$ as shown in Fig. 4.

%-------------------------------- f 4 --------------------------
\begin{figure}
\vspace{0.5cm}
\begin{center}
{\includegraphics[width=6cm]{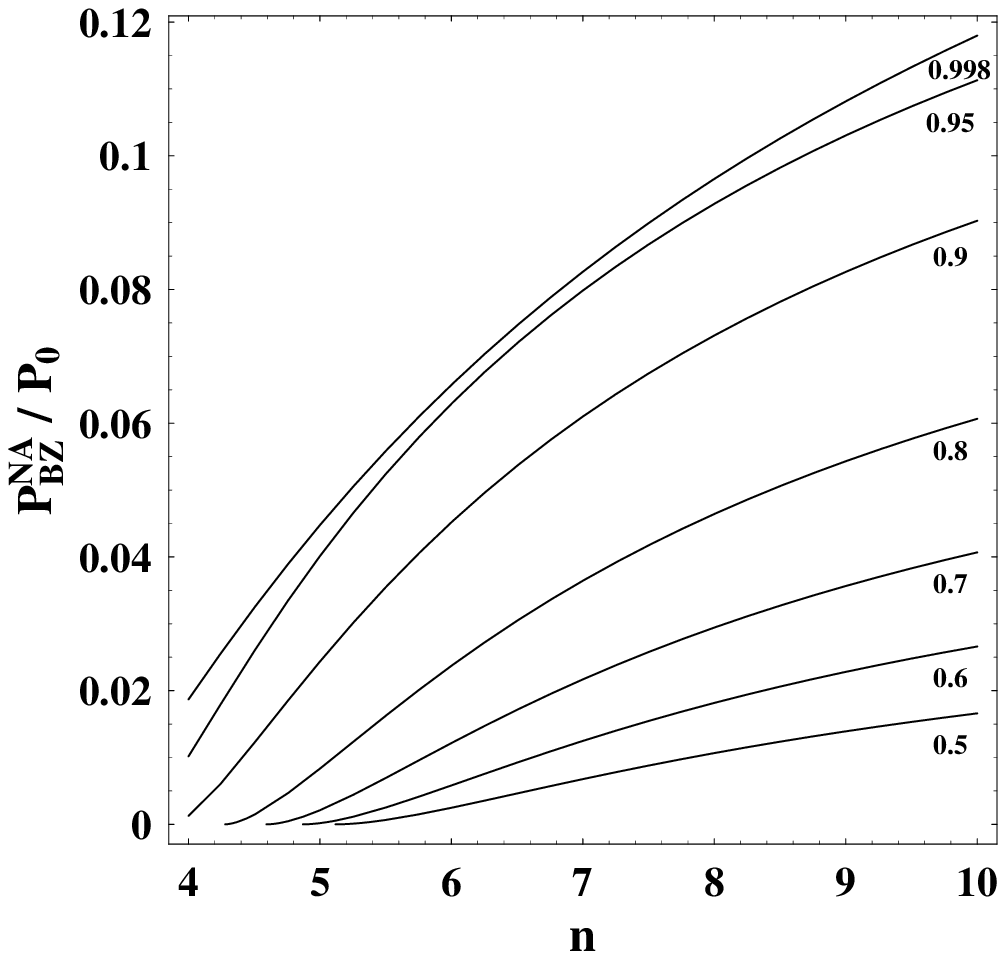}
 \centerline{\quad\quad\quad(a)}
 \includegraphics[width=6cm]{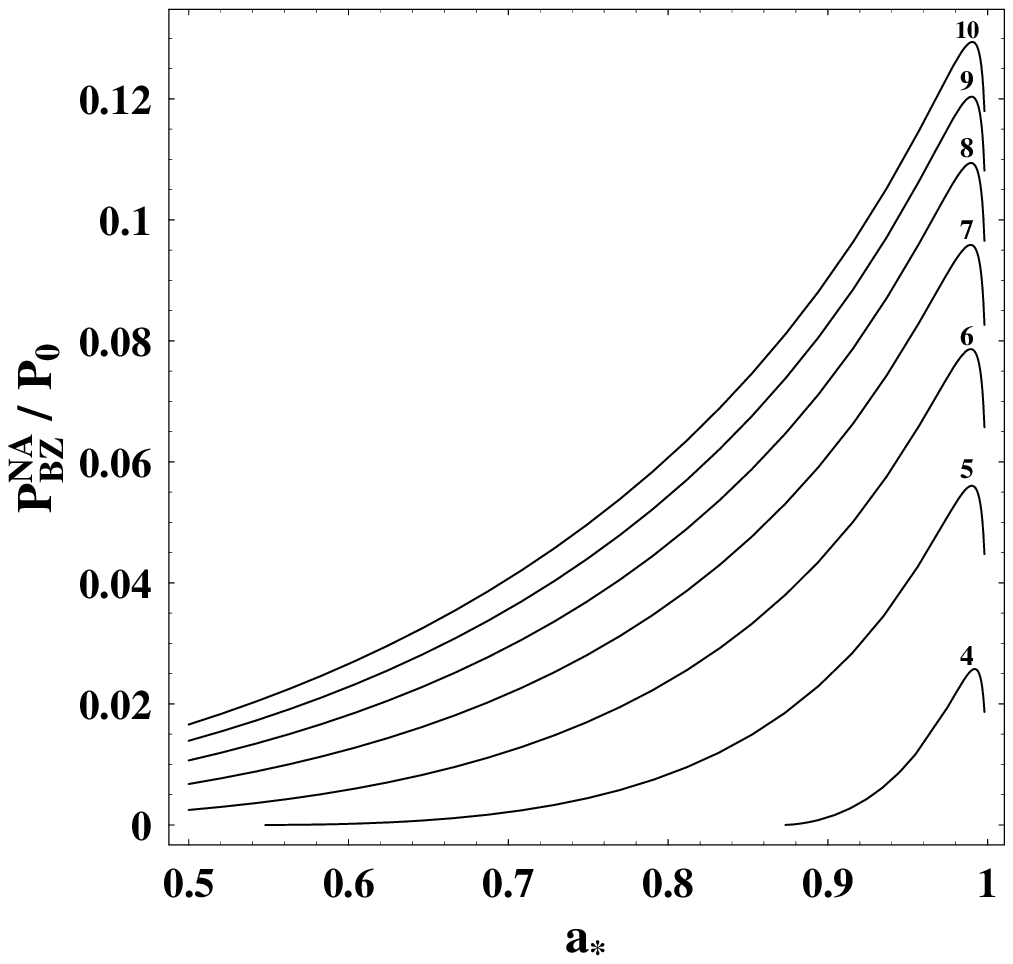}
 \centerline{\quad\quad\quad(b)}}
 \caption{The BZ power $P_{BZ}^{NA} $ (a)
varying with $n$ with the given values of $a_ * $, (b) varying
with $a_ * $ with the given values of $n$.}\label{fig4}
\end{center}
\end{figure}

%-------------------------------- sub 4 --------------------------

\subsection{3:2 ratio and electric current on disc}

As shown in Table 1, a very large power-law index $n$ varying from
6.17 to 6.29 is required by the 3:2 ratio of HFQPOs in the BH
binaries. It implies that the poloidal magnetic field increases
inward in a very steep way governed by equation (\ref{eq2}). It is
worthy to point out that this result is consistent with that in
W03a, where the power-law index $n$ varying from 6 to 8 is
required to explain a very steep emissivity (4-5) observed in
Seyfert 1 galaxy MCG-6-30-15.

The poloidal magnetic field arises most probably from the toroidal
electric current on the disc, and the current density $j_\varphi $
is related to $B_D^p $ by Ampere's law as follows.

%-------------------------------- eq 33 --------------------------

\begin{equation}
\label{eq33} j_\varphi = \frac{1}{4\pi }\frac{dB_D^p }{dr} =
\frac{1}{4\pi r_{ms} }\frac{dB_D^p }{d\xi } = - \frac{n\left(
{B_D^p } \right)_{ms} }{4\pi r_{ms} }\xi ^{ - \left( {n + 1}
\right)}.
\end{equation}

\noindent Inspecting equation (\ref{eq33}), we obtain two features
of the toroidal electric current on the disc.

(i) The minus sign in equation (\ref{eq33}) shows that the current
flows in the opposite direction to the disc matter; (ii) the
power-law index $n + 1$ in equation (\ref{eq33}) implies that the
profile of the current density in the central disc is steeper than
that of the magnetic field. We give a primary explanation for
these features as follows.

Probably the direction of the current arises from the effect of
the closed magnetic field lines on the charged particles of the
disc matter. In our model the angular velocity of the BH is
required to be greater than that of the disc to produce the inner
and outer hotspots on the disc. Therefore the charged particles of
the disc matter will be dragged forward by the closed field lines.
Considering that the mass of the positive charged particle, such
as proton, is much greater than that of electron, we think that
the bulk flow of electrons might be a little ahead of that of
protons. Probably the minute difference in the two kinds of the
bulk flows results in a current in the opposite direction to the
disc.

On the other hand, the flux of angular momentum transferred from
the rotating BH to the disc is given by

%-------------------------------- eq 34 --------------------------

\begin{equation}
\label{eq34} H_{MC}^{NA} = \frac{\partial {T_{MC}^{NA} }
\mathord{\left/ {\vphantom {{T_{MC}^{NA} } {\partial r}}} \right.
\kern-\nulldelimiterspace} {\partial r}}{4\pi r}.
\end{equation}

\noindent Substituting equations (\ref{eq8}) and (\ref{eq12}) into
equation (\ref{eq34}), we have

%-------------------------------- eq 35 --------------------------

\begin{equation}
\label{eq35} \begin{array}{l} H_{MC}^{NA} = \frac{1}{4\pi \xi
r_{ms}^2 }\left( {{\partial \theta } \mathord{\left/ {\vphantom
{{\partial \theta } {\partial \xi }}} \right.
\kern-\nulldelimiterspace} {\partial \xi }} \right)\left(
{\partial {T_{MC}^{NA} } \mathord{\left/ {\vphantom {{T_{MC}^{NA}
} {\partial \theta }}} \right. \kern-\nulldelimiterspace}
{\partial \theta }} \right)
\\ \quad\quad\quad = - \frac{\lambda a_ * \left( {1 +
q} \right)}{\pi r_{ms}^2 }\frac{\left( {1 - \beta } \right)\xi ^{
- 1}\mbox{G}\left( {a_ * ;\xi ,n} \right)}{\left[ {2\csc ^2\theta
- \left( {1 - q} \right)} \right]}.
\end{array}
\end{equation}

\noindent Incorporating equation (\ref{eq2}) with equation
(\ref{eq35}), we have

%-------------------------------- eq 36 --------------------------

\begin{equation}
\label{eq36} H_{MC}^{NA} \propto \xi ^{ - n}.
\end{equation}

Considering that the flux of angular momentum, $H_{MC}^{NA} $,
transferred to the disc will block the accreting particles, and
the blocking effect would be much more significant for electrons
than protons due to the mass differences. Thus more electrons
would stay in the region very close to ISCO. That is why the
current density consist of the bulk flow of electrons and it
varies with $\xi $ in a power law expressed by equation
(\ref{eq33}).

As a summary we intend to give a comment for the six parameters
involved in our model. The BH mass $m_{_{BH}} $ has its range
given by observations, and $B_4 \approx 10^5$ is required by the
hotspots for emitting X-ray. The parameters $\varepsilon $ and
$\delta $ are used to describe the non-axisymmetric magnetic field
on the BH horizon. In fact, neither the position nor the energy of
the inner and outer hotspots is sensitive to the values of
$\varepsilon $ and $\delta $, and we take $\delta = 0.5$ and
$\varepsilon = 0.2$ in calculations. Thus the rest two parameters,
$a_ * $ and $n$ are adjustable in fitting the 3:2 ratio, which
play very important roles in fitting the 3:2 ratio of HFQPOs.

In this paper a non-axisymmetric magnetic field on the BH horizon
is assumed in an ad hoc way. Unfortunately, we cannot give a good
explanation for its existence at present. Probably the
non-axisymmetric magnetic field on the BH horizon arises from
non-axisymmetric accretion disc, since the magnetic field is
brought and held by the surrounding magnetized disc. We hope to
approach this difficult task in future.\\

\noindent\textbf{Acknowledgments. }We thank the anonymous referee
for his/her many constructive suggestions. This work is supported
by the National Natural Science Foundation of China under Grant
Numbers 10173004, 10373006 and 10121503.

\end{document}